\documentstyle[12pt]{article}

\begin{document}
\setlength{\oddsidemargin}{20 mm}
\setlength{\topmargin}{-10 mm}
\setlength{\textwidth}{17.0cm}
\setlength{\textheight}{23.0cm}
\renewcommand {\baselinestretch}{1.5}
\baselineskip 18 pt plus 1pt minus 1pt
\setcounter{page}{0}
\begin{titlepage}
\vspace*{-2 cm}
\begin{flushright}CEBAF-TH-93-14\end{flushright}\vspace{2.5 cm}
\begin{center}
CHEMICAL RELAXATION TIMES IN A HADRON GAS AT FINITE TEMPERATURE
\vspace*{1.cm}\\
 J. L.  Goity \vspace*{5mm}\\
 {\it  Continuous Electron Beam Accelerator Facility\\
        Newport News, VA 23606, USA\\
and\\
Physics Department, Hampton University, Hampton, VA 23668}\vspace*{1. cm}\\
\end{center}
\begin{abstract}
The relaxation times of particle  numbers   in  hot hadronic
matter  with vanishing   baryon number are estimated using  the ideal gas
approximation and taking into account resonance decays
and annihilation processes  as the only  sources
of particle number fluctuations. Near the QCD critical temperature
 the longest relaxation times
turn out to be of the order of 10 fm  and grow
roughly exponentially to become of the order of $10^{3}$ fm at temperatures
around 100 MeV. As a consequence of such   long relaxation times,
a clear departure from chemical equilibrium
must be observed in the momentum distribution of secondary particles
produced in high energy nuclear collisions.

\end{abstract}
 \vspace*{.3 cm}
\end{titlepage}
\newpage
\setcounter{page}{1}
\section{Introduction}

Experimental studies of QCD   near and above the
critical temperature for the hadron to quark-gluon
phase transition are thought to be possible in ultrarelativistic
nuclear collisions. After such a collision takes place,
a hot spot of quark-gluon matter is expected to
remain at central rapidity. This secondary system
subsequently expands and cools  until
a phase transition to hadronic matter occurs. This first stage
of the evolution is the most relevant one for these studies.
Unfortunately, after the system enters  the hadronic phase,
thermal equilibrium is expected to persist, erasing strong interaction
signatures of this  first stage. The observed momentum distribution
of stable hadrons corresponds to the freeze-out configuration, expected to
correspond to temperatures well below the critical temperature.
 In present  high energy nuclear collisions
the final state of secondary particles shows
as many as five hundred particles, mostly
 pions, stemming from a source whose radius is about 5 fm., as determined
from pion interferometry.
Under these  conditions thermal equilibrium can persist
 down to temperatures
of the order of 100 MeV  $\cite{shuryak,g&l,gerber}$.  It is assumed
that freeze-out occurs suddenly, leading to a  thermalized momentum
distribution of the final state.    The composition of the
hadronic gas at  freeze-out,  however, will show an excess of
stable particles if  chemical equilibrium
is lost in the course of the evolution. In fact, inelastic processes
which can produce fluctuations in the number of particles and
restore chemical equilibrium in this way   are overcome by the expansion
 rate well before the stage of  freeze-out is reached.
This has been suggested $\cite{g&l,gerber,goity,kataja}$
as a natural explanation
for the   enhancement of the
pion spectrum at low transverse-momentum observed in present
 high energy nucleus collisions
$\cite{experimental}$.

While  the thermalization characteristic time
can be well determined  by considering  elastic
$\pi-\pi$ scattering  as the only source of equilibration
$\cite{shuryak,g&l}$, the   determination of the characteristic
times associated with chemical equilibrium  involve a large number
and variety of inelastic processes. Using the available information
on inelastic processes, in this work  we furnish   estimates
for these relaxation times.

The time evolution of the particle number densities per degree
of freedom $N_{\alpha}$ of the
different particles composing a system is determined by the following
first order differential equation:
\begin{equation}
g_{\alpha}\,\dot{N}_{\alpha}=
\sum_{\delta n=1}^{\infty} \delta n \; [P_{\alpha}
(\delta n) - P_{\alpha}
(-\delta n) ],
\end{equation}
where $g_{\alpha}=(2 \,s_{\alpha}+1)(2 \,I_{\alpha}+1)$
counts the number of degrees of freedom associated with
the spin and the isospin of the particle and
 $P_{\alpha}(\delta n)$
is the probability per unit
time and unit volume for the elementary inelastic processes
where the number of particles of type $\alpha$ change by $\delta n$.
$P_{\alpha}$ is given by the following expression:
\begin{eqnarray}
P_{\alpha}(\delta n)\!&=&\!\!
\sum_{ \{i\},\; \{f\} }\;S(\{i\})\,S(\{f\})\;\int \prod_{\beta_{i} \in \{i\}
}\;
 \prod_{\beta_{f} \in \{f\} } \; d\tilde{k}_{\beta_{i}}\;
 d\tilde{k}_{\beta_{f}}\;(2\pi)^{4} \,\delta^{4}(K_{i}-K_{f})\nonumber\\
&\times &\left|T_{i,f}\right|^{2} \;
\sigma_{\beta_{i}}(k_{\beta_{i}})\;\tilde{\sigma}_{\beta_{f}}(k_{\beta_{f}}),
\end{eqnarray}
where $\{i\}$ and $\{f\}$ denote respectively to the particle content
of the initial and
final states
in the elementary process. Obviously, these states must satisfy
the condition that $n_{\alpha}\{f\}-n_{\alpha}\{i\}= \delta n$.
Here $T_{i,f}$ is the scattering amplitude, $S(\{i\})$ and
 $S(\{f\})$ are the initial and final state statistical factors,
$K_{i}$ and $K_{f}$ are respectively  the
 momenta of the initial and final state,
 the Lorentz invariant
volume element in momentum space is as usual given by
\begin{equation}
d\tilde{k}_{\beta}\equiv \frac{d^{3}{k}_{\beta}}
{(2\,\pi)^{3} \;2\;E_{\beta}} ~~~~, \nonumber
\end{equation}
and the density factors, which take into account the statistics of the
particles, are given by:
\begin{eqnarray}
\sigma_{\beta}(k)&=&N_{\beta}(k) \\
 \tilde{\sigma}_{\beta}(k)&=&
\left\{ \begin{array}{ll}
1+N_{\beta}(k) & ,~~~~{\rm if ~\beta ~ is ~  a  ~ boson}\\
1-N_{\beta}(k)
&,~~~~{\rm  if ~  \beta  ~ is ~  a  ~ fermion}
\end{array}  \right.
\nonumber
\end{eqnarray}
where $N_{\beta}(k)$ is the number density per degree of freedom
of particles
of type $\beta$ with momentum $k$. In thermal equilibrium
at temperature T this number density is given by:
\begin{eqnarray}
  N_{\beta}(k)=  \left\{ \begin{array}{ll}
{\rm exp}[- \epsilon_{\beta}(k)/2T]/
(2 \;\sinh[\epsilon_{\beta}(k)/2T)] &
 ,~~~~{\rm if ~  \beta  ~ is  ~ a ~  boson}\\
{\rm exp}[- \epsilon_{\beta}(k)/2T]/
[2 \;\cosh(\epsilon_{\beta}(k)/2T)] &,~~~~
 {\rm if  ~ \beta  ~ is ~  a  ~ fermion }\end{array} \right.
\end{eqnarray}
where
$\epsilon_{\beta}(k)\equiv E_{\beta}(k)-\mu_{\beta}$,
and $\mu_{\beta}$ is a chemical potential.
A non-vanishing chemical potential can be associated with
conserved   and   approximately
conserved particle numbers. By the latter one means
those numbers whose characteristic relaxation times
are much greater than the thermal relaxation time
of the system.
A chemical potential be ascribed even to short lived resonances,
 since
resonances are constantly regenerated in the collisions
of stable states, and consequently, their lifetime
is not directly related to the   relaxation time
of their chemical potential.

{}From eqs. (2, 4, 5), and in the limit of small
chemical potentials,   the following
form for $\dot{N}_{\alpha}$ is obtained:
\begin{equation}
g_{\alpha}\,\dot{N}_{\alpha}=- \frac{1}{2\,T} \sum_{\{i\}} \sum_{\{f\}}
\delta n_{\alpha}(i|f)\;\sum_{\gamma} \,\delta n_{\gamma}(i|f)
\;\mu_{\gamma}\;\tilde{\Gamma}(i,f;T),
\end{equation}
where
 $\gamma$ runs over all particle species.
The explicit expression for $\tilde{\Gamma}(i,f;T)$ is
\begin{equation}
\tilde{\Gamma}(i,f;T)=\! \prod_{\beta_{i} \in \{i\}}
\prod_{\beta_{f} \in \{f\}}\;
S(\{i\})\,S(\{f\})\;
\int\; d\nu_{\beta_{i}} \;
d\nu_{\beta_{f}} \;(2\,\pi)^{4}\,\delta^{4}(K_{i}-K_{f})
 \; \left| T_{i,f}\right|^{2},
\end{equation}
with the definition
\begin{eqnarray}
 d\nu_{\beta}&\equiv &\frac{d\tilde{k}_{\beta}}{ 2\, \chi_{\beta}}\\
\chi_{\beta}&=& \left\{ \begin{array}{ll}
\sinh (E_{\beta}/2T) & ,~~~~{\rm if ~ \beta ~  is ~  a  ~ boson}\\
\cosh (E_{\beta}/2T) &,~~~~ {\rm if ~  \beta ~  is ~  a ~  fermion}\\
\end{array} \right. . \nonumber
\end{eqnarray}
For small chemical potential, we can write
\begin{equation}
\mu_{\beta}=T\;\frac{\Delta N_{\beta}}{N_{\beta}^{0}}=
T\; \frac{(N_{\beta}-N_{\beta}^{0})}{N_{\beta}^{0}},
\end{equation}
where  $N_{\beta}^{0}$ is the density at vanishing chemical potential.
 Since
we are considering the time evolution at constant temperature,
 $\dot{N_{\beta}^{0}}=0$, and eqn.(6) becomes a system of
first order linear differential equations with constant
coefficients.
 The characteristic times of this system of equations
are the chemical relaxation times we need.

In the case of baryons,   particle and antiparticle are taken
into account separately,
 while mesons are  considered as their own antiparticles by
including all
the members of each flavor multiplet. We are assuming
that there is no net baryon number, and therefore the chemical
potentials associated to a baryon and its antiparticle are  equal.

In a realistic situation, eqn.(6) requires a large body of information
about the multiple inelastic strong interaction processes.
Only partial information is available from data: for instance,
there are sufficient data on  resonance decays and $N-\bar{N}$ annihilation.
On the other hand, data on resonance production in hadronic
collisions at low energy
are poor, and   more exotic inelastic processes
are experimentally inaccessible.  As a consequence, approximations are
unavoidable. These approximations will consist in including
in eqn. (6) only those processes for which data are already available
and those whose rates can be estimated reliably enough.

For the sake of illustration, we first discuss the chemical relaxation time
in a  pion gas. In this case the
relevant inelastic process is $\pi\pi \leftrightarrow \pi \pi\pi\pi$,
for which an estimate  within Chiral Perturbation Theory is possible.
Later on, we consider a more realistic situation, where
all hadronic levels, except those containing  strange quarks, are included.
The main conclusion
is that chemical relaxation times are large, even at temperatures
close to the phase transition, supporting the conviction that in
a high energy heavy ion collision the evolution of the fire-ball
in the hadronic phase will proceed away from chemical equilibrium.

\section{Pion gas}

At low temperatures and in the absence of net baryon number,
 a gas of pions gives a
good approximation to the thermodynamics of
 a hadronic gas. As the temperature increases, heavier states become
 increasingly populated and eventually dominate the energy density
$\cite{gerber&l}$. Concerning the kinetic  properties of the hadronic gas,
it was found that the thermal relaxation time is reliably
obtained within the pion gas approximation if the temperature
is less than 130 MeV $\cite{schenk}$. On the other hand, as we find in this
work,
the relaxation times associated with chemical equilibrium
are poorly represented in this approximation if the temperature
exceeds 70 MeV. Since the question of chemical equilibrium
is turns out to be important
 at temperatures close to the critical temperature,
for practical purposes the pion gas  turns out to give
an unrealistic picture. For the sake of making this claim
clear, and also giving a simple illustration of the topic of this
work, we discuss the pion gas case in detail.

We  work in
the dilute gas approximation $(E_{\pi}>>2T)$, where
according to eqn.(6) the  time evolution of $\bar{N}_{\pi}\equiv
N_{\pi^{0}}+N_{\pi^{+}}+N_{\pi^{-}}$ becomes determined by
\begin{eqnarray}
  \frac{\partial}{\partial t}{\Delta \bar{N}_{\pi}}&\simeq&
\! -\frac{1}{12} \frac{\Delta \bar{N}_{\pi}}{\bar{N}_{\pi}^{0}}
\;\int\;\prod_{i=1}^{6} \; d\tilde{k}_{i} \; (2\,\pi)^{4}
\,\delta^{4}(k_1+k_2-k_3-k_4-k_5-k_6) \nonumber\\
&\times &\!{\rm exp}(-(E_{\pi}(k_1)+E_{\pi}(k_2)/T)
\sum_{{\rm isospin}}  \left|
 T(k_1\,k_2\rightarrow k_3\,k_4\,k_5\,k_6)\right|^{2}
\end{eqnarray}
Performing the integrations over the final state phase space and
the angular integrations in the initial state phase space, this equation
takes the following form:
\begin{eqnarray}
 \dot{\Delta\bar{N}_{\pi}}&\simeq &
-\frac{1}{4 \pi^4}
\frac{\Delta \bar{N}_{\pi}}{\bar{N}_{\pi}^{0}} \;
\int_{M_{\pi}}^{\infty} dE_1 \;\sqrt{E_1^2-M_{\pi}^{2}}\;
\int_{E_{2}^{min}}^{\infty} dE_2 \;\sqrt{E_2^2-M_{\pi}^{2}}\;\\
& \times &{\rm exp}(-(E_{1}+E_{2})/T)\;
\sqrt{s (s-4 M_{\pi}^{2})}\;
\sum_{  I_1,\,I_2} \sigma_{\pi\pi\rightarrow
\pi\pi\pi\pi}(s,I_1,I_2)\nonumber
\end{eqnarray}
where:
\begin{equation}
E_{2}^{min}=\left\{ \begin{array}{ll}
7\;E_{1}-4 \sqrt{3} \sqrt{E_{1}^{2}-M_{\pi}^{2}} &, ~~~ {\rm if} ~~
E_{1}<7\;M_{\pi}\\
M_{\pi}  & ,~~~ {\rm if} ~~ E_{1}>7\;M_{\pi}
\end{array} \right.
\end{equation}

The cross section is estimated by using the amplitude
obtained from the lowest order chiral Lagrangian. A
     practical simplification   consists in neglecting
the tree level contribution to the amplitude resulting from
twice iterating  the chiral Lagrangian term containing the four-leg vertices.
We expect that this approximation will not significantly affect
the results.  In the chiral limit, and after summing over the initial state
isospins $I_{1}$ and $I_{2}$, the   inelastic cross section,
 is:
\begin{equation}
\sum_{isospin\;\; I_1,I_2} \sigma_{\pi\pi\rightarrow
\pi\pi\pi\pi}(s,I_1,I_2)=\frac{67}{2^{17}\;3^{4} \;\pi^{5}}
\;\;\frac{s^{3}}{F_{\pi}^{8}}~~~~~~~F_{\pi}=93~~{\rm MeV}
\end{equation}
In this limit, the integrations over the initial energies are
readily performed,
and the    chemical relaxation time becomes:
\begin{equation}
\left.\tau_{\pi}\right|_{M_{\pi}=0}\sim
4700 \;\;\frac{F_{\pi}^{8}}{T^{9}}
\end{equation}
For the physical value of the pion mass we perform the integration
numerically. Table I shows the results for the chemical relaxation time.
For comparison,
 the thermal relaxation time of the pion gas $\cite{g&l}$ is also
displayed.

As $s$ increases the lowest order chiral Lagrangian
fails to give a good representation of the cross section,
overestimating its value. For this reason, at temperatures above
150 MeV   our results underestimate
the  length  of the chemical relaxation time for the pure
pion gas.

\section{Hadronic Gas}

In a hadronic gas a large variety of inelastic elementary processes
can take place. These processes  tend to maintain or restore
chemical equilibrium as they can produce fluctuations of the
different particle numbers.
Due to the large multiplicity of hadronic states and the large
variety of inelastic processes, it is clear that  a precise
analysis is impossible. Data are available only for resonance decays
and $N - \bar{N}$ annihilation processes, and to a lesser
extent, for resonance production in $\pi - N$ and $N -  N$  collisions
at low energy.
For this reason, we will only take into account the following
processes: (a) resonance decays, e.g.  $\rho
\rightarrow \pi\pi$, $N^{\ast} \rightarrow N\pi$,
 $N^{\ast} \rightarrow N\pi\pi$, etc.; (b) baryon-antibaryon and
excited meson annihilation into pions, e.g.
 $\bar{N} N \rightarrow \pi...\pi$, $\omega
\omega \rightarrow \pi...\pi$, etc. The purely pionic processes
considered in the previous section turn out to give small
corrections
in this more realistic picture, and are therefore neglected.
 Strange hadrons are  also neglected.
We expect that by solely including  these  classes of
processes, a reasonable estimate
of the relevant chemical relaxation times can be obtained.

Within the dilute gas approximation to eqn.(6) one can then write down:

\begin{eqnarray}
\theta_{\alpha} g_{\alpha} \dot{N}_{\alpha}&=&
\sum_{\beta}A_{\alpha\beta} \;\mu_{\beta}~~~~~~,
\end{eqnarray}
where $\alpha,\;\beta=\pi,\;N,\; N_{i}^{\ast},\;\Delta_{j}$,
and, $\theta_{\alpha}=1(2)$ if $\alpha$ is a meson(baryon).
The matrix $A_{\alpha \beta}$ is symmetric, and
for the class of processes we take into account its matrix elements
are given by the following expressions:
\begin{eqnarray}
A_{\pi \pi}&=&-\frac{1}{T} \;\left(\sum_{\alpha=N_{i}^{\ast},\,\Delta_{j}}\;
\theta_{\alpha}\,N_{\alpha} \;g_{\alpha} \;\Gamma_{\alpha}\;
\langle \delta n_{\pi}^{2}\rangle_{\alpha}
\right.\nonumber\\
&+&\left. \sum_{\Delta_{i},\,\Delta_{j}} g_{\Delta_{i}}\;
g_{\Delta_{j}} \;
\langle \delta n_{\pi}^{2}\rangle_{\Delta_{i}\,\Delta_{j}}\;\,
\Omega_{\Delta_{i}\,\Delta_{j}}(T)
\right. \nonumber\\
&+&\left. \sum_{N_{i},\,\bar{N}_{j}}\;
g_{N_{i}}\;(g_{N_{j}}-\delta_{ij})\;
\langle \delta n_{\pi}^{2}\rangle_{N_{i}\,N_{j}}\;\,
\Omega_{N_{i}\,N_{j}}(T)\right) \nonumber\\
A_{\pi N}&=&-\frac{2}{T }\;\left(\sum_{N_{i}^{\ast}}\;N_{N_{i}^{\ast}}\;
g_{N_{i}^{\ast}}
\;\Gamma_{N_{i}^{\ast}}\;
\langle \delta n_{\pi}\rangle_{N_{i}^{\ast}}
\right.\nonumber\\
&-&\left. \sum_{N_{i}}\; g_{N}\,(g_{N_{i}}-\delta_{i1})\;
\langle \delta n_{\pi}\rangle_{N\,N_{i}}
\; \Omega_{N\,N_{i}}(T)\right)\nonumber\\
A_{\pi \Delta_{j}}&=&
\frac{1}{T}  \;\left(N_{\Delta_{j}}\;g_{\Delta_{j}}\;\Gamma_{\Delta_{j}}
\;\langle \delta n_{\pi}\rangle_{\Delta_{j}}
\right. \nonumber\\
&+& \left. g_{\Delta_{j}}\;\sum_{\Delta_{i}}\;g_{\Delta_{i}}\;
\langle \delta n_{\pi}\rangle_{\Delta_{i}\,\Delta_{j}}\;
 \frac{1}{2}\;\Omega_{\Delta_{i}\,\Delta_{j}}(T)\right)\nonumber\\
A_{\pi N^{\ast}_{i}} &=&
\frac{2}{T } \;\left(N_{N_{i}^{\ast}}\;
g_{N_{i}^{\ast}}
\;\Gamma_{N_{i}^{\ast}}\;
\langle \delta n_{\pi}\rangle_{N_{i}^{\ast}}
\right. \nonumber\\
&+& \left. g_{N_{i}^{\ast}}\;
\sum_{N_{j}}\;(g_{N_{j}}-\delta_{ij})\;
\langle \delta n_{\pi}\rangle_{N_{i}^{\ast}\,N_{j}}\;\,
\Omega_{N_{i}^{\ast}\,N_{j}}(T)\right)\nonumber\\
A_{N N}&=&-\frac{2}{T } \;\left(\sum_{N_{i}^{\ast}}\;N_{N_{i}^{\ast}}\;
g_{N_{i}^{\ast}}
\;\Gamma_{N_{i}^{\ast}}\;\right. \nonumber\\
&+& \left. g_{N}\;\sum_{N_{i}}\;(g_{N_{i}}-\delta_{i1})\,(1+\delta_{i1})\;\,
\Omega_{N\,N_{i}}(T)\right)\nonumber\\
A_{N \Delta_{j}} &=& 0\nonumber\\
A_{N N^{\ast}_{i}}&=&\frac{2}{T } \;\left(N_{N_{i}^{\ast}}\;g_{N_{i}^{\ast}}
\;\Gamma_{N_{i}^{\ast}}\;
- g_{N}\;g_{N_{i}^{\ast}}\;\Omega_{N\,N_{i}^{\ast}}(T)\right)\nonumber\\
A_{\Delta_{i} \Delta_{j}}&=&
-\frac{1}{T} \;\left(\delta_{ij} N_{\Delta_{i}}\;
g_{\Delta_{i}}\;\Gamma_{\Delta_{i}}
+ 4\;g_{\Delta_{i}}\;g_{\Delta_{j}}\;
\frac{1}{2}\;\Omega_{\Delta_{i}\,\Delta_{j}}(T)\right)\nonumber\\
A_{\Delta_{i} N^{\ast}_{j}}  &=& 0 \nonumber\\
A_{N^{\ast}_{i} N^{\ast}_{j}} &=&
-\frac{2}{T } \;\left(\delta_{ij}\;N_{i}^{\ast}\;g_{N_{i}^{\ast}}
\;\Gamma_{N_{i}^{\ast}} + g_{N_{i}^{\ast}}\;(g_{N_{j}^{\ast}}-\delta_{ij})
\;\Omega_{N_{i}^{\ast} \,N_{j}^{\ast}}(T)\right)
\end{eqnarray}
where the following definitions have been used:
$\langle \delta n_{\pi}\rangle_{\alpha}$ is the average number of pions
produced in the decay of the state $\alpha$,
$\langle \delta n_{\pi}\rangle_{\alpha\,\beta}$
is the average number of pions
produced in the annihilation of $\alpha$ with $\bar{\beta}$,
and analogously for the corresponding averages of the square of the number of
pions.
We  have assumed that the annihilation averages are roughly
energy-independent at low energy, so that  they can
be factored out. Here
$\Gamma_{\alpha}$ is the width of the corresponding resonance,  and
  $\Omega_{\alpha\,\beta}$ can be expressed in terms of the
total annihilation cross section as follows:
\begin{equation}
\Omega_{\alpha\,\beta}=
\int \;\frac{d^{3} k_{\alpha}\;d^{3}k_{\beta}}
{(2 \,\pi)^{6}}\;{\rm exp}(-(E_{\alpha}+E_{\beta})/T)\;
\mid v(k_{\alpha},k_{\beta})\mid\;\sigma_{\alpha \,\beta
\rightarrow {\rm pions}},
\end{equation}
where  $\mid v(k_{\alpha},k_{\beta})\mid$  the relative speed
in the rest frame of one of the colliding partners. For
 $N - \bar{N}$ annihilation the following approximation
holds $\cite{bebie}$:
\begin{equation}
\Omega_{N\,N}(T)\sim N_{N}^2 \;\frac{1}{M} (b+4\,a\, \sqrt{M\,T/\pi})
\end{equation}
where  M is the nucleon mass, $a\sim 17 $ mb and $b\sim 39$ mb.GeV.
We   adopt here the following approximation: for all
possible  annihilation
processes the form shown in  (18) is used, with the
densities replaced by the densities of the partners in the collision
and the nucleon mass replaced by the average mass. This is a
crude guess, which  is required by the lack of  experimental access
to resonance annihilation processess. The following further approximations
are performed: (a) the averages
$\langle \delta n_{\pi}\rangle_{\alpha\,\beta}$ and
$\langle \delta n_{\pi}^{2}\rangle_{\alpha\,\beta}$
are taken to be the same
for all annihilation processes and  (b) resonances whose width is larger
than 250 MeV are not included.

We first consider  a simplified case, where only pions, nucleons
and $\rho$ mesons are taken into account.  We find that the longest
relaxation time corresponds to a configuration where
 $\mu_{\pi}\sim \frac{1}{2}\,\mu_{\rho}$ and, in the limit where
$\langle \delta n_{\pi}^{2}\rangle_{\rm{annihil.}} \sim
\langle \delta n_{\pi} \rangle_{\rm{annihil.}}^{2}$, $\mu_{N}\sim
 \frac{1}{2}\langle \delta n_{\pi} \rangle_{\rm{annihil.}}\mu_{\pi}$.
  In reality,
 $\langle \delta n_{\pi}^{2}\rangle_{\rm{annihil.}} >
\langle \delta n_{\pi} \rangle_{\rm{annihil.}}^{2}$, and the latter relation
is substantially affected.   The next longest
relaxation time corresponds to a configuration where the
chemical potentials of pions and  $\rho$ mesons are in the same
relation as before, but the nucleon
chemical potential  is now similar to $\mu_{\pi}$  and of opposite sign.
This is interpreted as follows: The longest relaxation time
has to do with how fast the  bulk excess of particles can be
annihilated.     The second longest time
measures how fast an excess of nucleons plus antinucleons  is  annihilated.

When we include all hadronic levels with no strangeness content,
a hierarchy of relaxation times results. Only those times longer than
the time scale characterizing the evolution of the system are of relevance.
It turns out that the longest times are similar to those obtained in the
simplified case. Thus, the approach to chemical equilibrium
is not substantially altered by the addition of more states.
It however occurs that a larger number of configurations
will have relaxation times which are long on this time scale.
These configurations show different patterns for the
chemical potentials. The only qualitative feature shared by  these patterns
is that  at temperatures below 160 MeV the chemical potentials
of mesonic resonances are all similar to each other, and approximately
given by twice the pion chemical potential. The baryonic chemical potentials
are also similar to each other, and much larger than $\mu_{\pi}$, with
$\mu_{N_{i}} \sim \kappa(T)\,\mu_{\pi}$, where $\kappa(T)
\sim 10\times 2^{(T_{0}-T)/20 {\rm MeV}}$, $T_{0}\sim 160$ MeV.
 Above that
temperature the baryonic chemical potentials become of the
same order as $\mu_{\pi}$, and
the mesonic ones show   values much larger than $\mu_{\pi}$.
This   indicates that    the
approach to equilibrium of those configurations with the longest
relaxation times  are driven at high temperatures mainly by the baryons,
and at lower
temperatures by the meson resonances.

We have checked that by removing  some states  the longest relaxation times
are hardly affected. There are also  minor effects
in the composition of the gas corresponding to those
longest times. We also observe  modest changes when the
annihilation cross section is changed by a factor of two or so.
 This suggests that the estimates
given here  should not be significantly altered
by taking more precise
values for the annihilation cross sections and by
 including  other  inelastic processes ignored in our approximations.
Finally, comparison with the results for the pion gas shows the inadequacy
of the latter as an approximation (taking  into account that
the pion gas results obtained here are trustworthy
  below 150 MeV, as mentioned in section 2).

One can draw the following picture of a hadronic gas which
drives out of chemical equilibrium in the course of the
expansion. Suppose that the system is large enough
so that  initially   equilibrium prevails. As the expansion rate
increases and becomes larger than the longest relaxation time
at the current temperature, a chemical potential associated to
the configuration of the corresponding eigenvector starts to develop.
As the expansion proceeds, it eventually becomes larger
than the second longest relaxation time, and  an additional
chemical potential starts to develop.  Thus,
in the course of the expansion, at each stage the
thermodynamic state of the hadronic gas will
be characterized by T, and as many chemical potentials as
the number of characteristic relaxation times which are
  longer than the   current expansion rate. The evolution of these
potentials
in the course of the expansion can in principle be
estimated.
In practice,   the typical radii of the
highest multiplicity systems  produced in  high energy nuclear collisions
are of the order of 10 fm at freeze-out. Thus, in this situation,
the relaxation times  ought to be compared with
 the size of the system rather than with the inverse expansion rate,
as this is smaller.
 Clearly,  chemical equilibrium cannot be sustained even
in the early stages after hadronization  occurs.

\vspace{15mm}
{\bf Acknowledgement:}
I thank the Theory Division of CERN for their kind hospitality and
  financial support
during the completion of part of this work. I also thank
Simon Capstick for carefully reading the manuscript.

 \newpage

\begin{center}
TABLES
\end{center}
\vspace*{15 mm}
\vspace{15mm}
\begin{center}
\begin{tabular}{|c|c|c|c|} \cline{1-4}
& & & \\
T& $\tau_{\pi} (M_{\pi}=0)$ &  $\tau_{\pi}(M_{\pi}=138 \;{\rm MeV})$
 & $\tau_{{\rm thermal}}$
 \\
 MeV & fm & fm  &fm     \\ \hline
100& 5150 &17000 & 17 \\  \hline
 120  & 1000    & 2300   &7 \\  \hline
 140  &  250   & 450   &3.3 \\  \hline
  160 &  75   &  120  &1.7 \\  \hline
 180  & 26    & 40   & 0.9 \\  \hline
 200  &  10   & 15   & 0.5 \\  \hline
\end{tabular}\vspace*{6 mm}\\
{\large{Table I}}

\end{center}
\vspace*{6 mm}\\
\vspace{15mm}
\begin{center}
\begin{tabular}{|c|c|c|} \cline{1-3}
&                   &           \\
T       & $\tau^{(A)} $ & $\tau^{(B)}$\\
MeV    & fm  &fm       \\ \hline
100     &$ 2200;\; 1200$  &$1060;\;470;\;20;\;8$      \\
       & $ 1300;\; 700$  &$ 1030;\;280;\;20;\;8 $   \\  \hline
120     & $ 260;\; 130$ & $165;\;65;\;15;\;8 $     \\
        & $ 200 ;\; 60$ &$ 140;\; 40;\; 15; \;8 $       \\  \hline
140     & $ 65;\; 25$   &$ 50;\;17;\;9;\;7    $    \\
        & $ 50 ;\;13$  &$ 35; \;15;\; 8;\; 5.5 $      \\  \hline
160     & $ 25;\;9 $  &$ 22;\;10;\;8;\;4     $    \\
        & $ 17;\; 4$  &$ 12;\; 9;\; 8;\; 4   $       \\  \hline
180     & $ 10;\;4 $  &$  13;\;9;\;7;\;4     $    \\
        & $ 7 ;\;2$  & $ 9.5 ;\; 7; \;5;\; 4 $         \\  \hline
200     & $ 5 ;\;2$  & $ 11;\;9;\;6;\;4    $     \\
        & $ 4 ;\;1$ & $ 9; \;7 ; \;4;\;3   $       \\  \hline
\end{tabular}\vspace*{6 mm}\\
{\large{Table II}}

\end{center}
\newpage

\begin{center}
CAPTIONS
\end{center}
\vspace*{15 mm}

\parbox{4.5in}{{\bf Table I:} {\it Results for
chemical relaxation time  $\tau_{\pi}$ in a pure
pionic gas. The last column shows the thermal relaxation time.}}
\vspace{15mm}

\parbox{4.5in}{{\bf Table II:} {\it Results for  chemical relaxation times
in a hadronic gas. (A) refers to the case where only pions, nucleons
and $\rho$-mesons are included. The two longest relaxation
times are given. (B) corresponds to the
inclusion of all relevant hadronic levels. The four longest
relaxation times are given.
The upper and lower rows    correspond repectively
to the cases $\langle \delta n_{\pi}^{2}\rangle_{\rm{annihil.} }=30$
and $40$. In all cases  $\langle \delta n_{\pi}\rangle_{\rm{annihil.}}=5$.}}

\end{document}